# Vacancy ordering and electronic structure of γ-Fe$_2$O$_3$ (maghemite): a theoretical investigation

**Ricardo Grau-Crespo,* Asmaa Y. Al-Baitai, Iman Saadoune and Nora H. De Leeuw**

Department of Chemistry, University College London, 20 Gordon Street, London, United Kingdom, WC1H 0AJ.

Email: r.grau-crespo@ucl.ac.uk

**Abstract.** The crystal structure of the iron oxide γ-Fe$_2$O$_3$ is usually reported in either the cubic system (space group P4$_3$32) with partial Fe vacancy disorder or in the tetragonal system (space group P4$_1$2$_1$2) with full site ordering and $c/a$≈3. Using a supercell of the cubic structure, we obtain the spectrum of energies of all the ordered configurations which contribute to the partially disordered P4$_3$32 cubic structure. Our results show that the configuration with space group P4$_1$2$_1$2 is indeed much more stable than the others, and that this stability arises from a favourable electrostatic contribution, as this configuration exhibits the maximum possible homogeneity in the distribution of iron cations and vacancies. Maghemite is therefore expected to be fully ordered in equilibrium, and deviations from this behaviour should be associated with metastable growth, extended anti-site defects and surface effects in the case of small nanoparticles. The confirmation of the ordered tetragonal structure allows us to investigate the electronic structure of the material using density functional theory (DFT) calculations. The inclusion of a Hubbard (DFT+U) correction allows the calculation of a band gap in good agreement with experiment. The value of the gap is dependent on the electron spin, which is the basis for the spin-filtering properties of maghemite.

1.  **Introduction**

Maghemite ($\gamma$-$Fe_2O_3$) is the second most stable polymorph of iron oxide. Contrasting with antiferromagnetic hematite ($\alpha$- $Fe_2O_3$), maghemite exhibits ferrimagnetic ordering with a net magnetic moment (2.5 $\mu_B$ per formula unit) and high Néel temperature (~950 K), which together with its chemical stability and low cost led to its wide application as magnetic pigment in electronic recording media since the late 1940's [1]. Maghemite nanoparticles are also widely used in biomedicine, because their magnetism allows manipulation with external fields, while they are biocompatible and potentially non-toxic to humans [2, 3]. Another promising application is in the field of spintronics, where it has been suggested that $\gamma$-$Fe_2O_3$ can be used as a magnetic tunnelling-barrier for room-temperature spin-filter devices [4, 5].

Maghemite occurs naturally in soils as a weathering product of magnetite ($Fe_3O_4$), to which it is structurally related [6]. Both maghemite and magnetite exhibit a spinel crystal structure, but while the latter contains both $Fe^{2+}$ and $Fe^{3+}$ cations, in maghemite all the iron cations are in trivalent state, and the charge neutrality of the cell is guaranteed by the presence of cation vacancies. The unit cell of magnetite can be represented as $(Fe^{3+})_8[Fe^{2.5+}]_{16}O_{32}$, where the brackets () and [] designate tetrahedral and octahedral sites, respectively, corresponding to 8a and 16d Wyckoff positions in space group Fd3m. The maghemite structure can be obtained by creating 8/3 vacancies out of the 24 Fe sites in the cubic unit cell of magnetite. These vacancies are known to be located in the octahedral sites [7] and therefore the structure of maghemite can be approximated as a cubic unit cell with composition $(Fe^{3+})_8[Fe^{3+}_{5/6} \square_{1/6}]_{16}O_{32}$.

The nature and degree of ordering of the iron vacancies in the octahedral sites has been the subject of investigations for several decades. If the cation vacancies were randomly distributed over the octahedral sites, as it was initially assumed, the space group would be Fd3m like in magnetite [8, 9]. The first indication of a departure from the Fd3m symmetry was reported by Haul and Schoon [10], who noticed extra reflections in the powder diffraction pattern of maghemite prepared by oxidising magnetite. Braun [11] later noticed that maghemite exhibits the same superstructure as lithium ferrite ($LiFe_5O_8$), which is also a spinel with unit cell composition $(Fe^{3+})_8[Fe^{3+}_{3/4}Li^{1+}_{1/4}]_{16}O_{32}$, and suggested this was due to similar ordering in both compounds. In the space group $P4_332$ of lithium ferrite, there are two types of octahedral sites, one with multiplicity 12 in the unit cell, and one with multiplicity 4, which is the one occupied by Li. In maghemite, the same symmetry exists if the Fe vacancies are constrained to these Wyckoff 4b sites, instead of being distributed over *all* the 16 octahedral sites. It should be noted, however, that some level of disorder persists in this structure, as the 4b sites have fractional (1/3) iron occupancies.

Oosterhout and Rooijmans [12] first suggested a spinel tetragonal superstructure with $c/a=3$, where the Fe atoms are completely ordered. A neutron diffraction study by Greaves [13] confirmed a higher degree of ordering than the one implied by the cubic $P4_332$ structure, and described this departure as a tetragonal distortion. The positions of the vacancies in the fully-ordered maghemite structure were obtained by Shmakov *et al.* [14] using synchrotron X-ray diffraction. This ordered maghemite structure has the tetragonal space group $P4_12_12$ with $a=8.347$ Å and $c=25.042$ Å (spinel cubic cell tripled along the *c* axis). The ion coordinates in the $P4_12_12$ structure have been recently refined by Jorgensen *et al.* based on synchrotron X-ray powder diffraction data [15].

Despite this progress in the structure determination of maghemite, the phenomenon of vacancy ordering in the lattice is not yet fully understood. It is not clear, for example, under which conditions, if any, vacancy disorder occurs. It has been suggested that the degree of ordering depends on crystal size, and that very small particles of maghemite do now show vacancy ordering [6, 16], although a recent investigation of needle-shaped maghemite nanoparticles with average size 240nm x 30nm has found the same tetragonal distortion with space group $P4_12_12$ as in the ordered crystal [17]. The thermodynamics of vacancy ordering in maghemite has not been investigated so far, in part because of the difficulty to control experimentally the level of ordering of the iron vacancies.

In this paper, we present a computational investigation of the energetics of vacancy ordering in maghemite. We will show that a fully ordered structure with tetragonal space group $P4_12_12$ is indeed the most stable configuration among all the possible ionic arrangements that are compatible with the partially disordered $P4_332$ structure, and that this stability arises from a most favourable electrostatic contribution. We then use this ordered structure to discuss the electronic properties of maghemite, which are relevant for potential applications of maghemite in the field of spintronics.

## 2. Computational details

The thermodynamics of ion disorder was investigated by the direct evaluation of the lattice energies of different ionic configurations, using interatomic potentials. This approach is based on the Born model of ionic solids [18], which assumes that the ions in the crystal interact via long-range electrostatic forces and short-range forces, including both the Pauli repulsion and dispersion attraction between neighbouring electron charge clouds. The short-range contribution has a simple analytical form, in this case a Buckingham potential, given by the expression:

$$V_{ij}(r_{ij}) = A_{ij} \exp(-r_{ij}/\rho_{ij}) - C_{ij}/r_{ij}^6 \qquad (1)$$

where $A_{ij}$, $\rho_{ij}$ and $C_{ij}$ are parameters specific to the interaction of the ions $i$ and $j$ and $r_{ij}$ is the ion separation. The electronic polarisability of the ions is included via the shell model of Dick and Overhauser [19], where each polarisable ion, here the oxygen ion, is represented by a core and a massless shell, connected by a spring. The polarisability of the model ion is then determined by the spring constant and the charges of the core and shell. The potential parameters used in this study were derived by Lewis and Catlow [20], and the calculations were performed with the GULP code [21-23].

The investigation of site-disordered structures using computer-modelling methods poses the problem of the large number of possible configurations that can exist for a particular supercell. We have used the methodology implemented in the program SOD (Site Occupancy Disorder [24]), which generates the complete configurational space for each composition of the supercell, and then extracts the subspace of symmetrically equivalent configurations. The criterion for the equivalence of two configurations is the existence of an isometric transformation that converts one configuration into the other and the transformations considered are simply the symmetry operators of the parent structure (the structure from which all configurations are derived via site substitution). This method typically reduces the size of the configurational space by one or two orders of magnitude, making the problem more tractable.

Although simulations based on classical interatomic potentials are known to perform very well in the description of ionic and semi-ionic solids, including iron oxides (e.g. [25-29]), quantum mechanical calculations are required when investigating their electronic and magnetic properties. We have performed electronic structure calculations, based on the density functional theory (DFT) in the generalized gradient approximation (GGA), using the Vienna Ab Initio Simulation Program (VASP) [30]. In order to improve the description of the Fe 3d orbitals, a Hubbard correction was added, using the so-called GGA+U methodology [31-34], in the approximation of Dudarev et al., (1998) where a single parameter, $U_{eff}$, determines an orbital-dependent correction to the DFT energy. In this work we have used $U_{eff}$ =4eV, which has been shown to provide a good description of the electronic and magnetic structures of different $Fe^{3+}$ oxides including hematite ($\alpha$-$Fe_2O_3$) [35, 36] and iron antimony oxide ($FeSbO_4$) [37-40]. The basis set size here is regulated by the cutoff energy ($E_{cut}$=400 eV in our study), in such a way that all plane waves with energies less than $E_{cut}$ are included. We have used frozen-core projector augmented wave (PAW) potentials [41], where the core consisted of orbitals up to and including $3p$ of Fe and $1s$ of O atoms. The Brillouin zone was sampled by a 3x3x1 Monkhorst-Pack mesh of k-points for the (1x1x3) supercell for the geometry optimisation, while a mesh of 6x6x2 was employed for the calculation of the electronic density of states. The calculations allow for spin polarisation of the wave functions, to reflect the magnetic character of maghemite. In maghemite, as in magnetite, the magnetic moments on the tetrahedral and octahedral sites are oriented in opposite directions, leading to ferrimagnetic behaviour [42], and we have therefore used the same magnetic configuration in our calculations.

## 3. Results and Discussion

*3.1 Configurational spectrum*

We first employ interatomic potential calculations to investigate the ordering of cation vacancies in γ-$Fe_2O_3$. Our starting point is the partially disordered cubic spinel structure with space group $P4_332$ initially suggested by Braun [11], where Fe ions and vacancies are distributed in the Wyckoff 4b octahedral positions. This structure is equivalent to lithium ferrite $LiFe_5O_8$, where the 4b positions are occupied by the Li cations. For this reason, we will call these positions "L" (for lithium) sites, even though we have no Li in the structures investigated in this work. An iron occupancy of 1/3 on the L sites makes the stoichiometry $Fe^L_{1/3}Fe_5O_8$. In the partially disordered cubic cell of maghemite, the 2.667 (or 8/3) iron vacancies are randomly distributed over the four L sites, together with 1.333 (or 4/3) iron cations. In order to have integer occupancies, we triple the unit cell along one axis (chosen to be *c*, to be consistent with the traditional convention for tetragonal systems). This 1x1x3 supercell thus contains 8 vacancies, which are now distributed, together with 4 iron cations, over the 12 L sites, and the coordinates of these positions for the 1x1x3 supercell are given in Table 1. Note that there are 12 layers, perpendicular to the <001> direction, containing octahedral sites with only one L-type site in each layer per simulation cell (Fig. 1).

The total number of combinations of the 4 Fe ions on the L sites of the supercell is 12!/(4! ×8!)=495, but only 29 of these are inequivalent, as determined using the SOD program. Table 2 lists the positions of the cations in each of the inequivalent configurations, together with their space groups, their degeneracies (how many times they are repeated in the full configurational space) and their relative energies, as calculated with the interatomic potential model. This information defines a multi-configurational model of vacancy ordering in maghemite, which is capable of describing the two extreme cases: if the energies of all the configurations are very similar, or differ very little compared with the thermal energy at the equilibration temperature, then the system is expected to be fully disordered. On the other hand, if one of the configurations is much more stable than the others, then the system should be ordered. A number of intermediate situations can also be described within the same framework, depending on the distribution of configuration energies.

The full configurational spectrum is shown in Fig. 2. Only one of these configurations has the space group $P4_12_12$, found by Shmakov *et al*. [14] for fully ordered maghemite. This configuration is indeed the most stable one, with a significant energetic separation from the second most stable configuration (32 kJ/mol). The energy range covered by the configurational spectrum is quite wide (~850 kJ/mol), indicating that full disorder is very unlikely. A more detailed analysis of the consequences of this energy spectrum will be given in section 3.3.

*3.2 The fully ordered maghemite structure: origin of its stabilisation*

The distinctive feature of the most stable configuration ($P4_12_12$) is the maximum possible homogeneity of iron cations and vacancies over the L sites. This configuration is the only one in which vacancies never occupy three consecutive layers; there are always two layers containing vacancies separated by a layer without vacancies, which instead contains $Fe^{3+}$ cations in the L sites (*e.g.* positions L1 - L4 - L7 - L10) and the $P4_12_12$ configuration is therefore the one that minimizes the electrostatic repulsion between these cations.

It is possible to check that the electrostatic interactions indeed dominate the relative stability of the different configurations over the whole spectrum: the total energies correlate well with the Coulomb-only energies obtained using formal charges for all ions (Fig. 3). The polarization of the anions is mainly responsible for the difference in the two energy scales, as polarization is known to compensate for the introduction of formal ionic charges in interatomic potential models [21]. Deviations from the straight line are mainly caused by relaxation effects, which are stronger for the least stable configurations. Based on this analysis, it is not surprising that the least stable configuration is the one with the maximum segregation of iron ions and vacancies over the L sites (iron cations in consecutive layers, e.g. L1 to L4, and vacancies in consecutive layers, L5 to L12), with an energy 847 kJ/mol above the $P4_12_12$ configuration.

The relaxed cell parameters for the ordered $P4_12_12$ structure are $a$=8.359 Å and $c$=24.854 Å. The ratio $c/3a$=0.991 shows a small but significant deviation from the cubic symmetry. In the paper by Shmakov *et al*. [14] no cell parameters are precisely given for the $P4_12_12$ structure, apart from stating that the cell is tripled along the *c* axis with respect to the original cubic structure (with $a$=8.347 Å). However, our result of $c/3a < 1$ is consistent with the observation by Greaves from neutron diffraction, that the tetragonal distortion accompanying vacancy ordering in maghemite slightly shrinks the crystal along the *c* axis with respect to *a* [13]. The bulk modulus obtained from our calculation of the ordered structure (211 GPa) is also in good agreement with the experimental value of Jiang *et al.* (203 GPa) [43].

Finally, we should note that, besides the ordered structure described here, there is another possible distribution of vacancies that gives the same $P4_12_12$ symmetry. This distribution, which is not listed in Table 2 as a configuration because it is partially disordered, can be described as follows. In the $P4_12_12$ space group, the L sites are divided into two symmetrically distinct positions, one with four-fold degeneracy, and the other with eight-fold degeneracy. While the ordered structure described above corresponds to full iron occupancy of the fourfold position, the distribution with half occupancy of the eightfold position also leads to $P4_12_12$ symmetry. However, we will show below that our calculated energetic spectrum of configurations strongly supports the full order scenario.

*3.3 Thermodynamics of ordering from canonical statistical mechanics*

In order to interpret the energy differences in the configurational spectrum in terms of the degree of vacancy ordering in the solid, we can estimate the probability of occurrence of each independent configuration *m* as a function of its energy $E_m$, its degeneracy $\Omega_m$ and the equilibration temperature *T*, using Boltzmann's statistics [24, 44, 45]:

$$P_m = \frac{\Omega_m}{Z} \exp(-E_m / k_B T) \qquad (2)$$

where $k_B = 8.314 \times 10^{-3}$ kJ/mol K is Boltzmann's constant, and

$$Z = \sum_m \Omega_m \exp(-E_m / k_B T) \qquad (3)$$

is the canonical partition function, which ensures that the sum of probabilities equals one. Figure 4 shows the probabilities of the most stable configuration ($P4_12_12$) and of the second most stable configuration (with space group $C222_1$) as a function of temperature. At 500 K, a typical synthesis temperature for maghemite [14], the cumulative probabilities of all the configurations excluding the most stable $P4_12_12$ is less than 0.1%. This contribution increases slowly with temperature, but at 800 K this cumulative probability, which measures the expected level of vacancy disorder, is still less than 2%. At temperatures above 700-800 K maghemite transforms irreversibly to hematite (α-$Fe_2O_3$), and considering higher temperatures is therefore irrelevant. It thus seems clear that perfect crystals of maghemite in configurational equilibrium should have a fully ordered distribution of cation vacancies.

It is important to realize that, in real samples, several factors can prevent this ordering to develop completely. First, synthesis temperatures are typically too low and preparation times too short to allow complete equilibration of the ionic configurations during the synthesis. Second, the nature of the ordering in the structure means that disorder of anti-site type is expected to be abundant. For example, if the ordering sequence along the *c* axis (two layers including vacancies plus one layer including iron cations in the L sites) is locally broken every few unit cells (for example, leading to two layers with iron cations in the L sites separated by only with vacancy layer), the overall symmetry of the crystal is not retained and the sample will appear disordered to diffraction methods. Third, in very small particles, surface effects might also alter the preferential distribution of cations, which could contribute to the absence of ordering reported in some maghemite nanoparticles [6, 16] .

*3.4 Electronic structure and magnetism*

The existence of a well-defined ordered structure of maghemite has an important advantage for the theoretical investigation of this material, as traditional electronic structure calculation techniques with periodic boundary conditions can be applied to the investigation of the electronic properties. In this section we discuss the electronic structure of ordered maghemite based on the results of our density functional theory (DFT) calculations.

As discussed in the methodology section, the ferrimagnetic ordering of the cation was forced in the calculation by assigning initial magnetic moments with opposite directions to $Fe^{3+}$ cations in octahedral and tetrahedral sites. However, the total magnetization of the $(Fe)_{24}[Fe]_{40}O_{96}$ cell was allowed to relax, and after self-consistency it reached the value of 80.0 $\mu_B$. This is the expected value if all the iron cations are in high-spin state, and corresponds to 2.5 $\mu_B$ per $(Fe)_{3/4}[Fe]_{5/4}O_3$ formula unit, in agreement with experimental measurements [1]. The magnetic moments are well localized on the Fe ions, and the integration of the spin density inside PAW spheres around the cations yields 4.19 $\mu_B$ for the octahedral Fe ions at L sites, 4.16±0.01 $\mu_B$ for the other octahedral Fe ions, and -4.03±0.01 $\mu_B$ for the tetrahedral cations. There are small net moments (0.14 $\mu_B$ or less) on the O anions. The iron magnetic moments obtained by Greaves [13] from neutron diffraction were 4.41 $\mu_B$ for octahedral and -4.18 $\mu_B$ for tetrahedral sites, in good agreement with our results.

Figure 5 shows the total electronic density of states and its projections over the 3*d* orbitals of octahedral and tetrahedral iron cations, and the 2*p* orbitals of the O anions. The top of the valence band is mainly of O 2*p* character, while the occupied 3*d* levels of Fe lie around 6-7 eV below the Fermi level. The bottom of the conduction band is mainly populated by the unoccupied 3*d* levels of (octahedrally coordinated) Fe. Therefore, maghemite is a charge-transfer type of insulator, and the first excitation term should correspond to the transfer of electrons from the $O^{2-}$ anions to the octahedral $Fe^{3+}$ cations.

The calculated band gap of around 2 eV is in agreement with experiment (2.0 eV according to Ref. [46]). However, the DOS around the Fermi level is not symmetric, and there is a difference in band gap between majority spin (2.2 eV) and minority spin (1.8 eV) electrons. The top of the valence band is higher in energy for majority spin electrons while the bottom of the conduction band is lower in energy for minority spin electrons. This is an important feature of the electronic structure of maghemite, as it is related to its potential use as a magnetic tunnelling-barrier for spin-filter devices. In these devices, the spin of the current electrons is controlled by an insulator film with an exchange splitting in the conduction band, through which tunnelling occurs preferentially for one of the spin components [47]. The potential of maghemite for this application has been previously suggested by other authors [4, 5].

**4. Conclusions**

This work represents the first attempt to investigate the phenomenon of vacancy ordering in γ-$Fe_2O_3$ (maghemite) from an energetic point of view. Our results show clearly that full vacancy ordering, in a pattern with space group $P4_12_12$, is the thermodynamically preferred situation in the bulk material. This stability arises from a minimal Coulombic repulsion between $Fe^{3+}$ cations for this configuration. However, deviation from perfect order can be expected because the low-temperature formation of maghemite does not guarantee an equilibrium growth of the crystals. Also, the presence of anti-site type disorder and surface effects in nanocrystals could contribute to deviation from the ideal ordering of the vacancies. We have also shown that maghemite is a charge-transfer type insulator with a spin-dependent band gap, which suggests its suitability for applications in spintronics.

**Acknowledgments**

This work made use of the facilities of HECToR, the UK's national high-performance computing service, via our membership of the UK's HPC Materials Chemistry Consortium, which is funded by EPSRC (EP/F067496). We acknowledge support from the EU-funded Marie Curie research and training network MIN-GRO (grant MRTN-CT-2006-035488). A.Y. A-B. is grateful to UCL for an Overseas Research Studentship (ORS) Award and the Iraqi government for funding.

**Table 1.**

**Coordinates of the L sites in the calculation supercell. These positions corresponds to the Wyckoff 4b sites of cubic space group P4$_3$32, expanded to a 1x1x3 supercell.**

| Position Label | Coordinates | | |
|---|---|---|---|
| | $x$ | $y$ | $z$ |
| L1 | 7/8 | 3/8 | 1/24 |
| L2 | 1/8 | 7/8 | 3/24 |
| L3 | 5/8 | 5/8 | 5/24 |
| L4 | 3/8 | 1/8 | 7/24 |
| L5 | 7/8 | 3/8 | 9/24 |
| L6 | 1/8 | 7/8 | 11/24 |
| L7 | 5/8 | 5/8 | 13/24 |
| L8 | 3/8 | 1/8 | 15/24 |
| L9 | 7/8 | 3/8 | 17/24 |
| L10 | 1/8 | 7/8 | 19/24 |
| L11 | 5/8 | 5/8 | 21/24 |
| L12 | 3/8 | 1/8 | 23/24 |

**Table 2.**

**Fully ordered configurations in the 1x1x3 supercell. The labels of the iron positions in the L sites follow the convention given in Table 1. Energies are given with respect to the lowest energy configuration.**

| Iron positions | Degeneracy | Space group | $\Delta E/kJ.mol^{-1}$ |
|---|---|---|---|
| L1, L4, L7, L10 | 3 | $P4_12_12$ | 0 |
| L1, L3, L7, L9 | 6 | $C222_1$ | 32 |
| L1, L3, L7, L10 | 24 | P1 | 53 |
| L1, L5, L6, L10 | 12 | $P2_1$ | 77 |
| L1, L5, L6, L8 | 12 | C2 | 87 |
| L1, L3, L7, L11 | 12 | C2 | 106 |
| L1, L2, L5, L9 | 24 | P1 | 116 |
| L1, L3, L7, L8 | 24 | P1 | 136 |
| L1, L2, L5, L8 | 24 | P1 | 149 |
| L1, L2, L7, L8 | 6 | $P2_12_12_1$ | 167 |
| L1, L3, L6, L8 | 12 | $P2_1$ | 182 |
| L1, L2, L5, L10 | 12 | $P2_1$ | 213 |
| L1, L3, L5, L10 | 12 | P1 | 215 |
| L1, L2, L6, L7 | 12 | C2 | 235 |
| L1, L3, L7, L12 | 24 | P1 | 276 |
| L1, L3, L6, L7 | 24 | P1 | 280 |
| L1, L4, L5, L6 | 24 | P1 | 310 |
| L1, L3, L5, L7 | 12 | C2 | 343 |
| L1, L3, L4, L10 | 24 | P1 | 380 |
| L1, L3, L4, L7 | 24 | P1 | 413 |
| L1, L2, L5, L6 | 12 | $P2_1$ | 425 |
| L1, L2, L3, L8 | 12 | C2 | 470 |
| L1, L2, L3, L7 | 24 | P1 | 501 |
| L1, L3, L5, L6 | 24 | P1 | 560 |
| L1, L2, L3, L6 | 24 | P1 | 608 |
| L1, L3, L4, L6 | 12 | $P2_1$ | 640 |
| L1, L2, L4, L5 | 12 | P1 | 652 |
| L1, L2, L3, L5 | 24 | P1 | 722 |
| L1, L2, L3, L4 | 12 | $P2_1$ | 847 |

**Figure captions:**

Figure 1: Possible positions for the iron vacancies in the 1x1x3 supercell. The 12 L sites, which should be populated with 4 iron ions and 8 vacancies, are marked as larger spheres.

Figure 2: Energetic spectrum of configurations for 4 iron ions and 8 vacancies distributed over the L sites in a 1x1x3 supercell of the cubic maghemite structure.

Figure 3: The relationship between the total lattice energies and the electrostatic energies of the different vacancy configurations.

Figure 4: Probabilities of the configurations as a function of temperature.

Figure 5: Electronic density of states corresponding to the ordered structure and its projection over Fe 3*d* and O 2*p* orbitals.

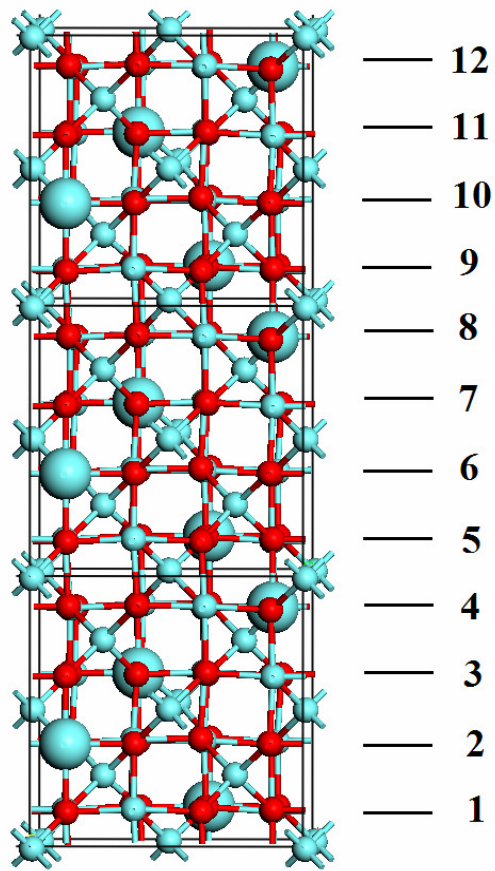

**Figure 1**

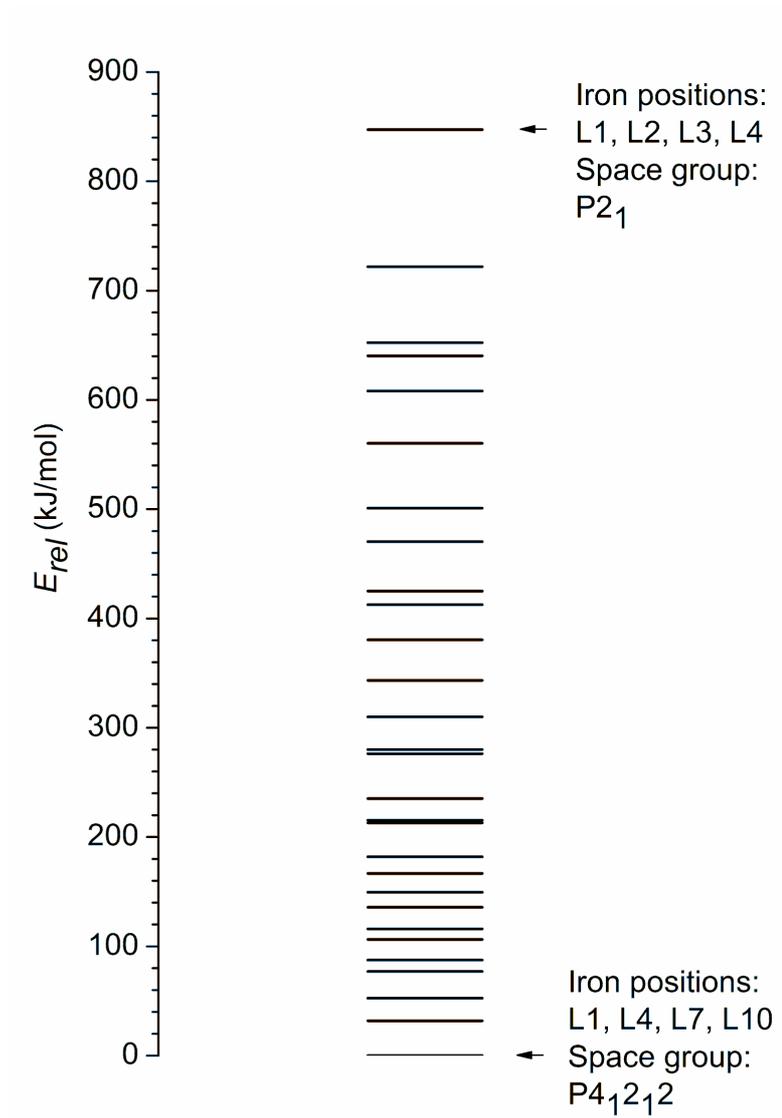

**Figure 2**

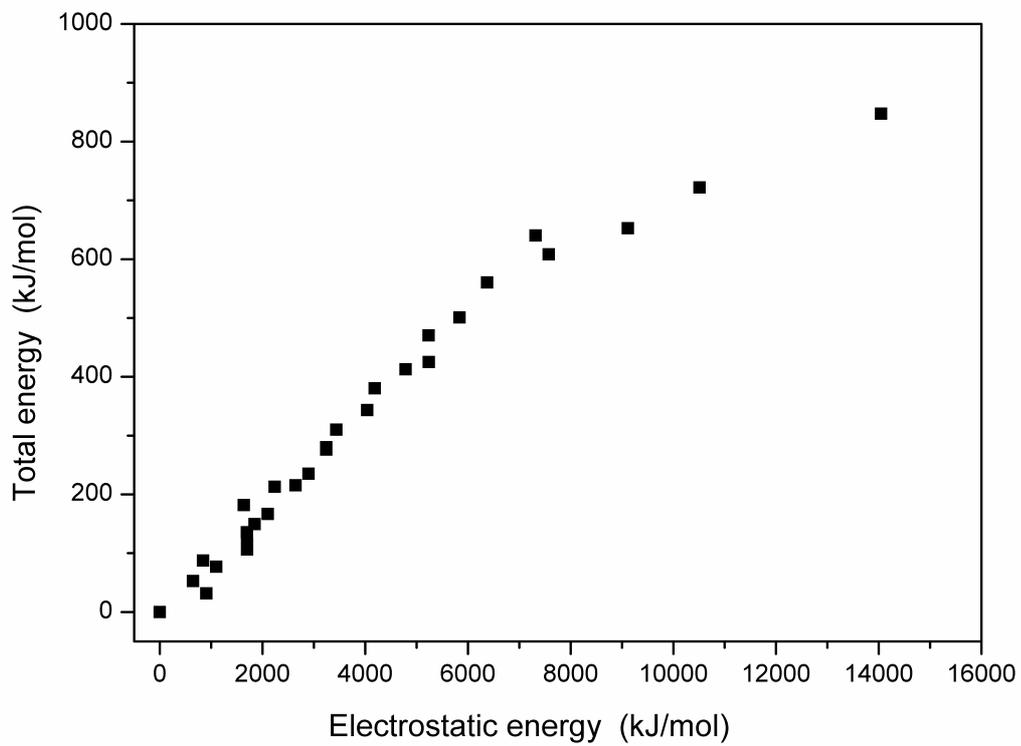

**Figure 3**

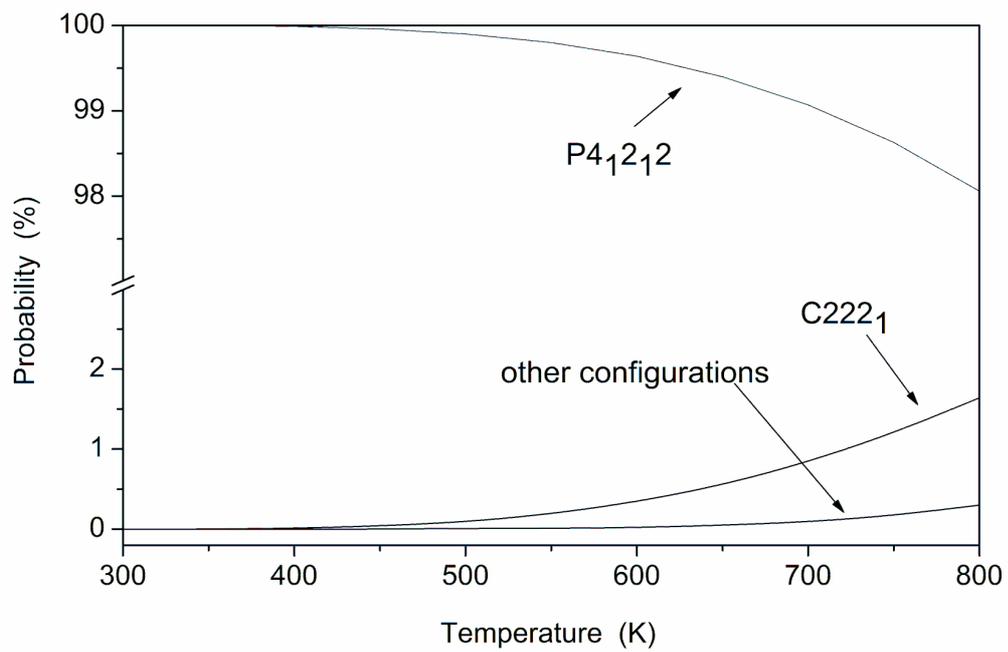

**Figure 4**

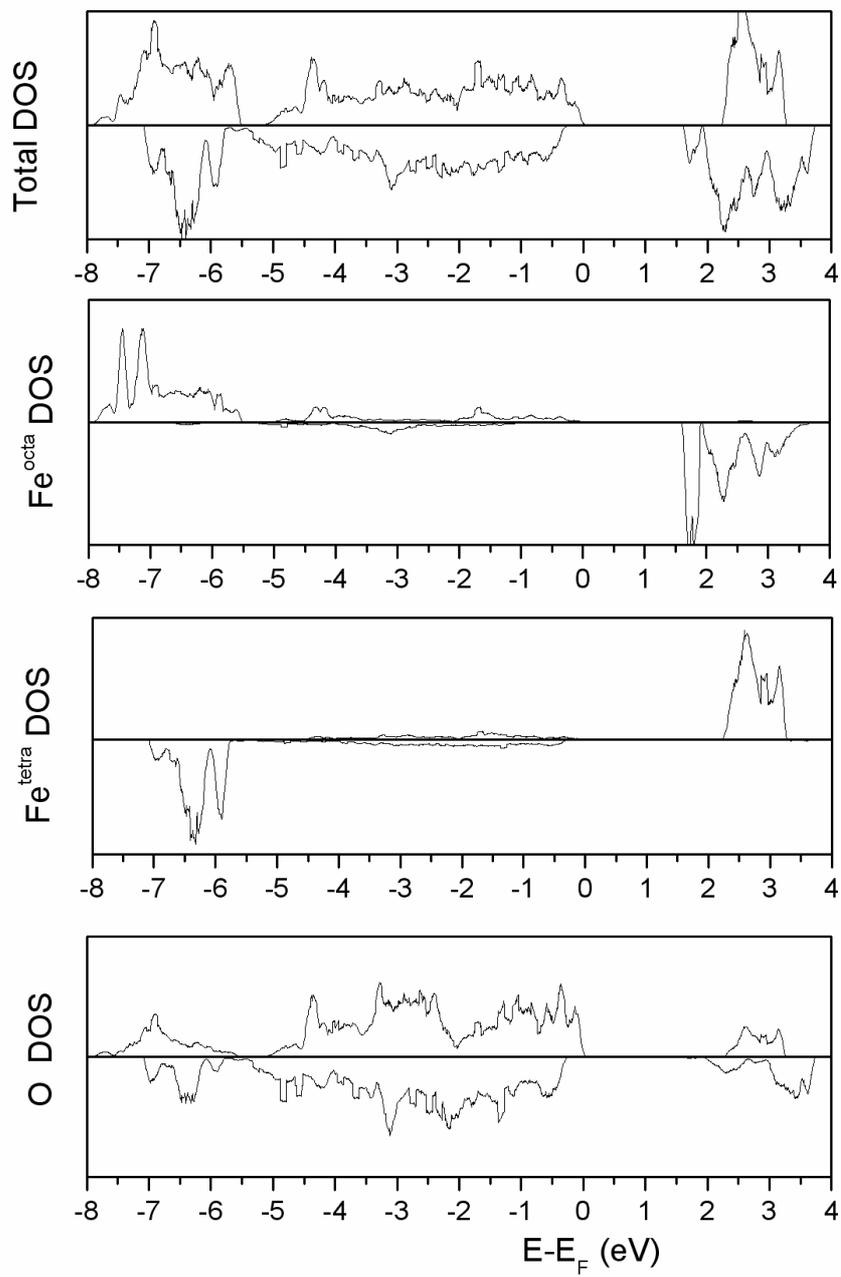

**Figure 5**